\documentclass[useAMS,usenatbib]{mn2e}

\usepackage{graphicx} 
\usepackage{comment} 
\newcommand{\oii}{[O\,{\sc ii}]}

\begin{document}
\title[CSPs and Abundances in the MW bulge and EGs]{Composite Stellar Populations and Element by Element Abundances in the Milky Way Bulge and Elliptical Galaxies}
\author[Baitian Tang, Guy Worthey, and A. Bianca Davis]{Baitian Tang$^{1}$\thanks{E-mail:
baitian.tang@email.wsu.edu (BT)} 
, Guy Worthey$^{1}$\thanks{E-mail:
gworthey@wsu.edu (GW)}, and A. Bianca Davis$^{2}$\\
$^{1}$Department of Physics and Astronomy, Washington State
University, Pullman, WA 99163-2814, USA\\
$^{2}$Department of Physics, The Ohio State University, 191 Woodruff Avenue, Columbus, OH 43210, USA}

\date{Accepted  . Received  ; in original form 2014}

\pagerange{\pageref{firstpage}--\pageref{lastpage}} \pubyear{ }

\maketitle

\label{firstpage}

\begin{abstract}
This paper explores the integrated-light characteristics of the Milky
Way (MW) bulge and to what extent they match those of elliptical
galaxies in the local universe. We model composite stellar populations
with realistic abundance distribution functions (ADFs), tracking the
trends of individual elements as a function of overall heavy element
abundance as actually observed in MW bulge stars.  The resultant
predictions for absorption feature strengths from the MW bulge mimic
elliptical galaxies better than solar neighborhood stars do, but the
MW bulge does not match elliptical galaxies, either. Comparing bulge
versus elliptical galaxies, Fe, Ti, and Mg trend about the same for
both but C, Na, and Ca seem irreconcilably different. 

Exploring the behavior of abundance compositeness leads to the
concepts of ``red lean'' where a narrower ADF appears more metal rich
than a wide one, and ``red spread'' where the spectral difference
between wide and narrow ADFs increases as the ADF peak is moved to
more metal-rich values. Tests on the systematics of recovering
abundance, abundance pattern, and age from composite stellar
populations using single stellar population models were performed. The
chemical abundance pattern was recovered adequately, though a few
minor systematic effects were uncovered. The prospects of measuring
the width of the ADF of an old stellar population were investigated
and seem bright using UV to IR photometry.
 
\end{abstract}


\section{Introduction}

For the better part of a century, the pursuit of chemical abundances
in astronomical objects has driven scientific innovation in the hope
of better understanding the origin and evolution of the universe and
the objects in it. Recently, this quest has extended to the integrated
light of early type galaxies, which seem to have
heavy element abundance ratio patterns that both do and do not fit
those seen in the Milky Way galaxy (MW), e.g., \citet{j12}.

Stellar abundances for stars in the MW that span a large range of
[Fe/H] show a pattern of heavy element enrichment that is largely
consistent with two sources for chemical enrichment, Type II and Type
Ia supernovae, where the ratio of Type Ia products increases with
increasing metallicity\footnote{Contrary to stellar spectroscopy
  tradition, we define ``metallicity'' as [M/H] $\approx$ log
  $Z/Z_\odot$, where $Z$ is the mass fraction of heavy elements. We
  assume in this work that a stellar evolutionary isochrone at fixed
  $Z$ will not vary with detailed chemical mixture, because isochrone
  sets that do have yet to be computed.}  \citep{wst1989}. Type II
supernovae, the core collapses and bounces of massive stars, are
thought to be rich in elements seeded by $^{12}$C plus the addition of
$^{4}$He nuclei, termed alpha-capture elements, or $\alpha$ elements
for short. They thus include even-numbered elements O, Ne, Mg, Si, S,
Ar, Ca, Ti, and possibly Cr, although Cr is more often included in the
group of elements termed the iron peak. Type Ia supernova, runaway
deflagration obliterations of white dwarfs, have a signature more
tilted toward the iron peak group, though some models produce
substantial quantities of Si, S, Ar, Ca, and Ti as well
\citep{n97}. Since the two sources are so disparate in origin, it is
not difficult to imagine many ways in which the relative proportions
could be made to shift and mix in different ways in different
environments.  Possible causal variables include the time interval
since star formation, the mass function at formation, the binary fraction, and
the heavy element composition.

There is now a plethora of evidence that this two-source picture is
too simple (e.g. \citealt{e93}) especially when stars from dwarf
spheroidal satellite galaxies are considered
(e.g. \citealt{geisler2005,shetrone2001,shetrone2003}) and also when
MW bulge stars are considered (e.g. \citealt{f07,b13,m08}). To explain
the discrepancies, it is further hypothesized that the Type II
enrichment pattern may change with progenitor mass and progenitor
chemical abundance \citep{f07}.

More evidence of multiple sources of enrichment is found in the
integrated light of early type galaxies. The initial result that more
massive elliptical galaxies had higher [Mg/Fe] \citep{w92} was
satisfactorily explained as a Type II/Type Ia ratio effect. But
\citet{w98} considered more elements (Na, Ca, and N) and could not
reconcile the trends in MW disk, MW bulge, and elliptical galaxies
under a two-source model.

The MW bulge has the potential to be a good analog for elliptical
galaxies, being a spheroidal component of the Galaxy and having a
stellar age that predates most of the disk \citep{o95,z03}, and yet
being near enough so that individual stars can be studied in some
detail. In that spirit, \citet{tfw} made integrated light models by
integrating the observed bulge luminosity function of the giants,
predicting TiO strengths and $VJHK$ colors. Both bulge stars and local
stars were used as spectral templates. The fascinating conclusion of
comparisons of those models with elliptical galaxies is that the bulge
template matched better than the solar neighborhood template.

Using this conclusion as a springboard, we explore in this paper
firstly the idea that using bulge templates is a superior match to
early type galaxies using additional observables than \citet{tfw} and
secondly the hypothesis that elemental ratio changes are
the cause.

To our goals adequately, the issue of compositeness in the stellar
populations must be addressed. Compositeness is a term that in general
would encompass mixtures of stellar population ages and
abundances. However, in cases of systems that formed most of their
stars in the first half of the universe's existence, the issue of age
is strongly suppressed due to (1) the strong decline of stellar
population luminosity with increasing age \citep{bc93} combined with
(2) the approximate three-halves rule \citep{w94a} that states that,
for example, a factor of three youthening of a population can be
counterbalanced by a factor of two increase in metallicity and the
integrated spectrum will change very little. By virtue of the fact
that a factor of three in age looms large against the age of the
universe for an old population, but a factor of two is small in
comparison to the range of two to three orders of magnitude for
overall heavy element abundance, it follows that compositeness in age
is very minor in effect compared to compositeness in heavy element
abundance.

This paper is organized as follows: Composite stellar populations
(CSPs) with metallicity-dependent chemical 
composition are illustrated in $\S$ \ref{sect:cc}. After that, we
confront the models with observables from three elliptical samples in $\S$
\ref{sect:comp}. The implications are discussed in $\S$
\ref{sect:disc}, and then a brief summary of the
results is given in $\S$ \ref{sect:con}.

\section{Composite Stellar Populations with Metallicity-dependent
  Chemical Composition} 
\label{sect:cc}

An aspect of compositeness that might conceivably have been present in
\citet{tfw} is that stars of inappropriate heavy element abundance may
have been inserted into the luminosity function model.

\textbf{Models:} A version of integrated-light models \citep{w94a,t98}
that use a new grid of synthetic spectra in the optical \citep{lee09}
in order to investigate the effects of changing the detailed elemental
composition on an integrated spectrum was used to create synthetic
spectra at a variety of ages and metallicities for single-burst
stellar populations. 

For this work, we adopt the isochrones of \citet{b08,b09} using the
thermally-pulsing asymptotic giant branch (TP-AGB) treatment described
in \citet{mar08}. This treatment is calibrated by comparing with
AGB stars in the Magellanic Clouds. Perhaps due
to counting statistics \citep{frogel90,sf97,bc03,salaris14}, the
numbers of AGB stars might be over-predicted \citep{gir10,gir13} in
this model version. Indeed, \citet{tw13} found abruptly reddened $V-K$
for metal poor SSPs at the age of 0.1, 1 and 2 Gyr from this isochrone
set. For now, however, our main concern is the optical wavelength
region, so we are insulated from this effect. Furthermore, the models
are modular as regards isochrone libraries, and swapping from one set
to another does not affect our conclusions.

Following \citet{poole10}, stellar index fitting functions were
generated from indices measured from the stellar spectral libraries of
\citet{v04} and \citet{lickx}, both transformed to a common 200 km
s$^{-1}$ spectral resolution. Multivariate polynomial fitting was done
in five overlapping temperature swaths as a function of $\theta_{eff}$
= 5040/T$_{eff}$, log g, and [Fe/H]. The fits were combined into a
lookup table for final use. As in \citet{w94a}, an index was looked up
for each bin in the isochrone and decomposed into ``index'' and
``continuum'' fluxes, which added, then re-formed into an index
representing the final, integrated value after the summation. This
gives us empirical index values. After that, additive index deltas
were applied as computed from the grid of \citet{lee09} synthetic
spectra when variations in chemical composition are needed. The grid
of synthetic spectra is complete enough to predict nearly arbitrary
composition changes.

\textbf{Abundance errors:} Regarding error propagation from individual stars to integrated light,
the calibration of the indices depends directly upon high resolution
analysis in that the hundreds of local stars that are fit have stellar
atmospheric parameters and abundances taken from the body of previous
high resolution work. We therefore expect systematic drift (from that
source) approximately equal to the error in the mean. For argument's
sake, if the scatter is 0.2 dex, and the sample is 100 stars, the
systematic drift should be of order $\sigma_{sys} \approx 0.2 /
\sqrt{100} = 0.02$. True errors from this source will be smaller for
well populated parts of the HR diagram such as G dwarfs and K
giants. 

Mildly more serious is a systematic effect from the Milky Way itself
in that the local stellar [X/Fe]\footnote{``X'' represents any one of
  the heavy elements discussed in this work.} trends are frozen into
the index fits. If these are not scaled-solar then they introduce
systematic drifts. Examining results from local stars, however, and
concentrating on thin disk stars only,
e.g. \citep{chen00,bensby03,bensby14,reddy06}, the [X/Fe] trends
versus [Fe/H] appear flat within a 0.1 dex range for stars near solar
metallicity ([O/Fe] might have a stronger tilt than that, but it is
difficult to tell within the increased uncertainty of this element),
and scatterlings appear in substantial numbers if thick disk stars are
included. A unified high resolution study of the exact sample of stars
that enter the low-resolution index fits has not been done. We
therefore must presume that the stars fitted obey the average
trend. In integrated light, fortunately, these stars are weak-lined
and few in number, leading to a few-hundredths change in overall
[X/Fe] for unfavorable cases.  Overall, we expect that systematic
abundance errors are a few hundreths of a dex for most [X/Fe], but
higher for elements that are difficult to measure with high resolution
spectroscopy (C, N, and O, for example)
\citep{grevesse07,asplund09,ryde10}.

The most serious systematic error is a purely integrated-light problem
arising from modeling uncertainties in stellar temperatures along the
isochrone. Temperature changes masquerade as either age or [M/H]
drifts, rendering an absolute [M/H] quite uncertain. This does not,
however, affect [X/Fe] measurements except via subtle
amplification/attenuation effects if the [M/H] is chosen incorrectly.

\textbf{Other caveats:} Observed indices that the model grids simply
do not cover occur from time to time. Reasons are as follows. The
observational errors (in this work) are not dominated
by photon statistics or wavelength solution errors. Errors in matching
line of sight velocity dispersion are present, and might contribute
more strongly. But the dominant error in the observations is probably
relative fluxing, in the sense of incomplete removal of instrumental
signatures in the spectrophotometric shape over tens or hundreds of
angstroms of wavelength span. Evidence that this effect is serious can
be found in that indices with larger wavelength spans in their
definitions from blue side to red side often drift more from the
models than narrower ones. Also, this defect shows up more in weaker
features than in stronger ones.

On the model side, the index fitting functions rest upon stellar
spectra for which the fluxing is imperfect as well, and so the same
kinds of fluxing effects can be expected at a modest level from the
stellar index fitting process as well. The fitting process also runs
the risk of oversimplifying the behavior of the indices with stellar
parameters.

\subsection{Abundance Distribution Functions (ADFs)}
\label{sect:adf}

In the process of making composite stellar populations with
single-burst ages but composite abundance distribution functions
(ADFs) there is clear empirical guidance for the shape of the
function. It has been clear for many years that the simplest one-zone,
constant-yield, no inflow, no outflow, instantaneous-recycling model
of chemical evolution produces too many metal-poor stars compared to
observations in the solar neighborhood (e.g. \citealt{p1997}). It has
also become clear in recent years that other chemically-evolved places
in the universe also have ADFs that are more narrow than the Simple
model \citep{wdj96,w05}.

\begin{figure}
\centering
\includegraphics [scale=0.4]{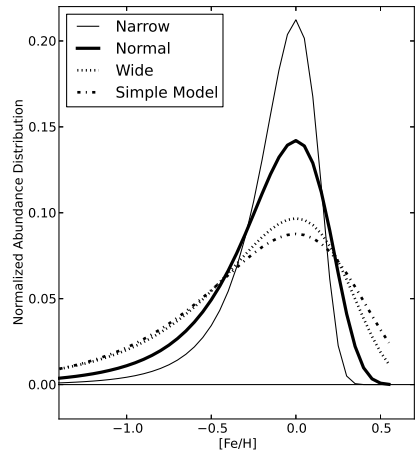} 
\caption{The four analytical ADFs we consider are
  shown, normalized to integral unity given bin widths of 0.1
  dex, and set to peak at solar abundance. The ADFs come in narrow
  (thin line), normal (thick line), and wide (dotted line) variants of
  the rational decreasing yield model, and the still-wider one zone
  Simple Model with yield $= 0$ (dash-dot line). The most metal rich population
  we can consider is [M/H] = 0.65 due to model
  limitations.}\label{fig:adfs}
\end{figure}

Figure \ref{fig:adfs} shows the Simple Model for the case of heavy
element yield equal to the solar value, along with the analytical
function variants that we adopt in this study. A convenient analytical
function is the rational decreasing yield formula from \citet{w05}. It
narrows the Simple model by having the yield start high and decrease
with increasing metallicity.  We use parameters $p=0.00019$ and
$\epsilon = 0.004$ in the formula for a smooth curve that has a FWHM
of 0.62 dex, and we call it our normal width ADF. From there, we scale
the width a factor of 1.5 narrower (narrow-width ADF), or a factor of
1.5 wider (wide-width ADF), and also transform it along the [M/H]
axis, as desired. The Simple Model is the widest of all, at a FWHM of
1.06 dex.

\begin{figure}
\centering
\includegraphics [scale=0.4]{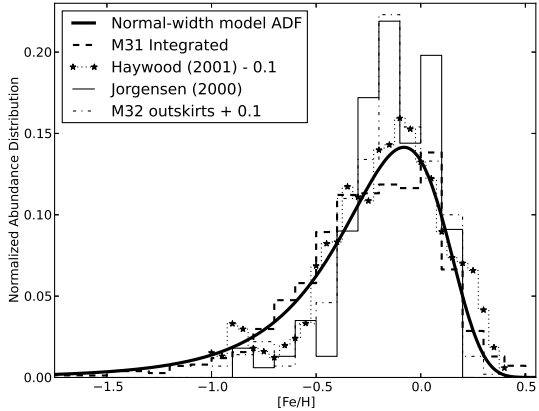} 
\caption{Observed ADFs are compared, normalized to integral one with a
  bin width of 0.1 dex along the [Fe/H] axis. Also shown is our normal
  width fitting function (heavy line), set to peak at [Fe/H]
  $=-0.1$. Other observed cases are M31, considered as a whole (dashed
  histogram) from photometry of red giants \citep{w05}), a field with
  mostly M32 stars (dash-dot histogram) also from photometry
  \citep{w2004} and shifted to the right by 0.1 dex for comparison,
  and two solar cylinder analyses, one from \citet{jorg2000} (thin
  solid histogram) and one from \citet{hay2001}, with corrections to
  the whole cylinder taken from \citet{wg1995} and shifted 0.1 dex to
  the left for comparison. Roughly, the observed ADFs are comparable
  in width to, or a bit narrower than, the analytical
  function.}\label{fig:adists}
\end{figure}
\begin{figure}
\centering
\includegraphics [scale=0.4]{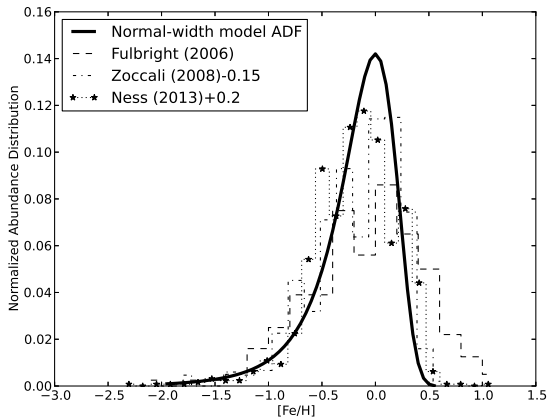} 
\caption{Milky Way Bulge ADFs from \citet{f06,z08,n13} (dashed,
  dash-dotted, and star-marked dotted lines, respectively) are shown
  in comparison with the normal-width analytical function (bold
  line). }\label{fig:buladf}
\end{figure}

While the rational decreasing yield model is not a complete physical
model in itself, it does a good job of reproducing the observations,
some of which are summarized in Figures \ref{fig:adists} and
\ref{fig:buladf}.  Note that some authors, e.g., \citet{hill11,babu10,
  n13}, argue that the MW bulge ADF is a multiple-peaked function that
plausibly represents distinct stellar populations. Without
contradicting that in the slightest, we pragmatically note that the
match between the observed and model ADFs is nevertheless fairly good,
especially in an integrated sense, and so we use the single-peaked
function throughout this paper.

\subsection{Milky Way Bulge Chemical Composition}
\label{sec:cc}

Our composite models are sensitive to element ratios, and the trends
of element ratios are included as a function of [M/H]. For
incorporation into the models, [Fe/H] and [X/Fe] are treated with
exactitude at the spectral library level, but at the isochrone level
[Fe/H] is equated with [M/H] due to the lack of elemental mixture
sensitivity in the isochrones. 

The MW bulge literature abundance measurements come from the following sources.

\begin{enumerate}
\item{\citet{m08} acquired high resolution near
    infrared spectra of 19 MW bulge giants. C, N, O and Fe
    abundances were obtained by spectrum synthesis of a number of lines
    from the MARCS \citep{g08} 1D hydrostatic model atmospheres.
}
\item{Using similar methods as \citet{m08}, \citet{ryde10} showed the
    C, N, and Fe abundances of  11 MW bulge giants. We exclude one
    outlier: Arp 4203, due to its unusual C and N abundances. } 
\item{\citet{a10} observed optical spectra of 25 Galactic bulge giants
    in Baade's window. O, Na, Mg, Al, Si, Ca, Ti and Fe abundances
    were derived from 1D local thermodynamics equilibrium analysis
    using Kurucz \citep{c97} and MARCS models. We choose the abundance
    ratios determined by MARCS models for consistence among different
    sets of measurements. These author's oxygen abundances are 
    excluded, because oxygen's trend is much different than the
    other samples at super-solar metallicity. 
    In addition, Alves-Brito et al. argue that the oxygen 
    abundances obtained from several IR OH lines (e.g., Mel\'endez sample) are
    preferable to these determined 
    from only one or two optical forbidden lines.
}
\item{\citet{b13} presented element abundance analysis of 58 dwarf
    and sub-giant stars in the MW bulge using microlensed
    spectra. MARCS model stellar atmospheres with Fe I NLTE (non-local
    thermodynamic equilibrium) corrections were employed to extract
    the stellar parameters. 
}
\item{\citet{hill11} performed observations upon a sample of stars in
    the Baade's window with GIRAFFE spectrograph at the VLT. Mg and Fe
  abundances were derived in 163 bulge clump giants using the MARCS models.
\item{\citet{johnson12} presented Na, Al, and Fe abundances of 39 red
    giant branch (RGB) stars and 23 potential inner disk red clump stars
    located in Plaut's window, while \citet{johnson13} showed the
    abundances of [Fe/H], [O/Fe], [Si/Fe], and [Ca/Fe] for 264
    RGB stars in three Galactic bulge off-axis fields. Note that these
    abundances are calculated relative to Arcturus. In order to
    consist with other sources which use solar abundances as
    reference, we adjust the
    reported abundances by adding the Arcturus abundances reported by Johnson et al. }
\item{\citet{rich05,rich07,rich12} collected a total of 61 bulge
  giants using NIRSPEC at the Keck telescope. They derived abundances
  for Fe, C, O and four $\alpha$ elements (Mg, Si, Ca, and Ti) with
  reference solar abundances from \citet{gs98}.}}
\end{enumerate}

To incorporate the empirical abundance trends into our
integrated-light models, we fit the observed [X/Fe]--[Fe/H]
relations. We use a one-error least square fitting,
in which the variances of [X/Fe] are considered as weights during the
procedure of finding the minimum $\chi^2$. We adopt an uncertainty
of 0.10 dex for the Mel\'endez sample and 0.11 dex, 0.09 dex for C and
N abundances in the Ryde sample. Most of the abundances in the other data sets
have individual errors, and the few without error estimates are excluded
before the fitting. For simplicity, we consider only two possible fitting functions, a
linear function or a quadratic function. The better one is
chosen after individual inspection for both.

\begin{figure*}
\hspace*{-0.7in}
\includegraphics [width=1.2\textwidth]{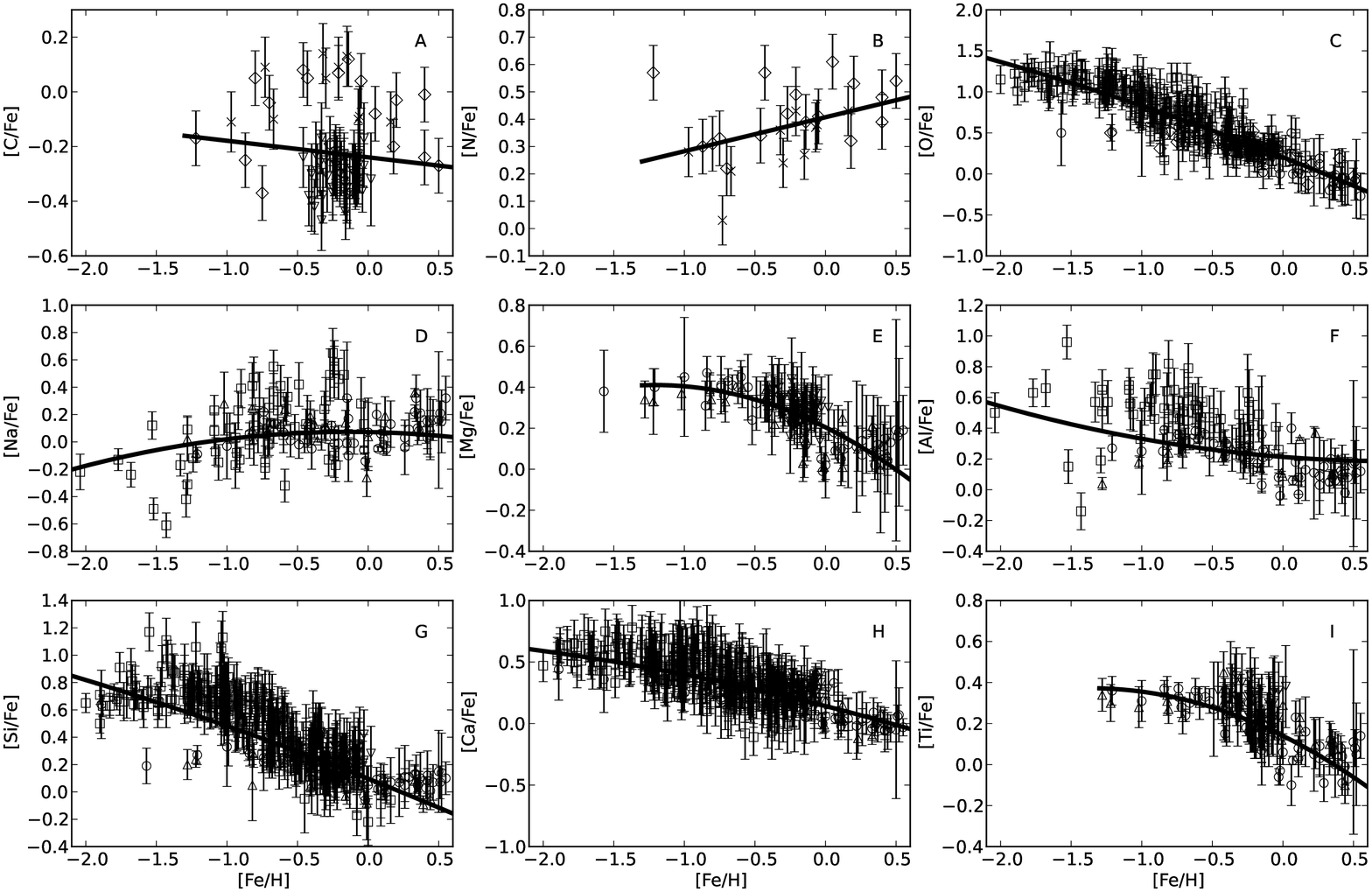} 
\caption{Abundance trends in the Milky Way bulge. Measurements from
  \citet{rich05, rich07, rich12} (open downward triangles),
  \citet{johnson12, johnson13} (open squares), \citet{hill11} (open
  pentagons), \citet{b13} (open circles), \citet{a10} (open
  upward triangles), \citet{ryde10} (crosses) and 
  \citet{m08} (open diamonds) are shown. Our fits are shown as thick solid
  lines. Errorbars are the published uncertainties.}\label{fig:elem}
\end{figure*}

\begin{itemize}
\item{C and N}

Most of the carbon abundances from \citet{m08,ryde10} are close to
solar, while \citeauthor{rich05} derived sub-solar abundances
[C/Fe]$\sim-0.2$. This disparity might originate from the complexity
of infrared spectral feature lines: C, O, and other stellar parameters
are obtained by simultaneous spectral fitting of several CO and OH
molecular bands, which are sensitive to temperature, gravity and
micro-turbulence \citep{origlia02}. In the cases of C and N, one might
be concerned with alterations in surface abundance due to dredge up of
nucleosynthetic processes. In the case of carbon, the mere fact that
carbon abundances are near-solar or even sub-solar imply that the
third dredge-up has not brought up significant amount of CNO-cycled
rest-products to the surface \citep{m08}.  However, super-solar
nitrogen abundances are observed and seem to require extra sources
besides the CNO cycle such as the neutrino process in massive stars
suggested by \citet{t95} or primary nitrogen. In this work, we model
[C/Fe] and [N/Fe] as linear functions of [Fe/H] (Figure
\ref{fig:elem}A and \ref{fig:elem}B).

\item{O and Mg}

According to the abundance yields of type II supernovae (e.g.:
\citealt{ww95}), O and Mg are two of the primary alpha elements
produced, and they are produced in almost the same ratio for stars of
disparate mass and progenitor heavy element abundance.  In Figure
\ref{fig:elem}C and \ref{fig:elem}E, O differs from Mg below
    [Fe/H]$\sim-1$, due to the scarcity of available data points for
    [Mg/Fe] at these metallicities. We fit them with quadratic
    functions.

\item{Na and Al}

Na and Al are closely affected by the excess neutrons associated with
$^{22}$Ne generation (the beta decay of $^{18}$F to $^{18}$O) during
the helium burning \citep{a71,handbook}. In Figures \ref{fig:elem}D
and \ref{fig:elem}F, [Na/Fe] is an increasing function of [Fe/H], while
[Al/Fe] decreases as metallicity increases. 
We fit them with quadratic functions.


\item{Si, Ca, and Ti}

Based on the similar trend of Si, Ca, and Ti as a function of [Fe/H],
we fit these three elements with quadratic functions (Figures \ref{fig:elem}G,
\ref{fig:elem}H, and \ref{fig:elem}I). These three elements are
grouped together as explosive alpha elements sharing a similar
nucleosynthetic origin. However, they do not show the same trend in
elliptical galaxies \citep{w11,w14,c14}. We search for the physical
reason underlying this disorder of facts in $\S$ \ref{sect:disc}.

\end{itemize}

Generally speaking, the overall trend is clear in all the panels of
Figure \ref{fig:elem}, but the scarcity of data for [Fe/H] $<-1.0$
renders the detailed form of the function vague for some elements.
When integrated, the fraction of metal poor stars is small in elliptical
galaxies, so the uncertainty of fitting function at [Fe/H] $<-1.0$ does
not affect our conclusions.

\subsection{300 km s$^{-1}$ Elliptical Galaxy Chemical Composition}

Since elliptical galaxies also seem to have a unique chemical
composition but we do not have the luxury of star-by-star chemical
analysis, we assume abundance ratios that are constant at all [M/H].
We incorporate the most up-to-date elliptical galaxy abundance ratios
into our CSP models, and call that variant the EG CSP. \citet{j12,w14}
and \citet{c14} have obtained the abundances of several major elements
from the integrated light spectra. Given the disparities of
their methods, it gives confidence that the final results are
generally consistent with each other (See \citealt{c14}). In this
work, we adopt the set of element abundance ratios extracted from
stacked 300 km s$^{-1}$ SDSS early-type galaxy spectra, namely,
[C/Fe]$=$0.21, [N/Fe]$=$0.27, [O/Fe]$=$0.28, [Na/Fe]$=$0.43, [Mg/Fe]$=$0.22,
[Al/Fe]$=$0.0, [Si/Fe]$=$0.16, [Ca/Fe]$=$0.02, [Ti/Fe]$=$0.12, set to be
constant at all [M/H].

\section{Results}
\label{sect:comp}

In this section, we show several indices for the three CSPs that we
assemble: 1. CSPs with scaled solar element abundances (SS CSPs, solid
lines); 2. CSPs with Galactic bulge element abundances (GB CSPs,
dashed lines); 3. CSPs with 300 km $s^{-1}$ elliptical galaxy element
abundances (EG CSPs, dotted lines).  We compute the CSPs at the age
of $2, 4, 6, 8, 10, 12$, and $14$ Gyr with five shifted [M/H]
distributions: peak [M/H]$= {-0.4, -0.2, 0.0, 0.2, 0.4}$.  We present
model indices at $\sigma=300$ km s$^{-1}$ resolution, where the
transformation from 200 km s$^{-1}$ to 300 km s$^{-1}$ is
accomplished via smoothing of synthetic model spectra.  

\begin{figure*}
\hspace*{-0.7in}
\includegraphics [width=1.2\textwidth]{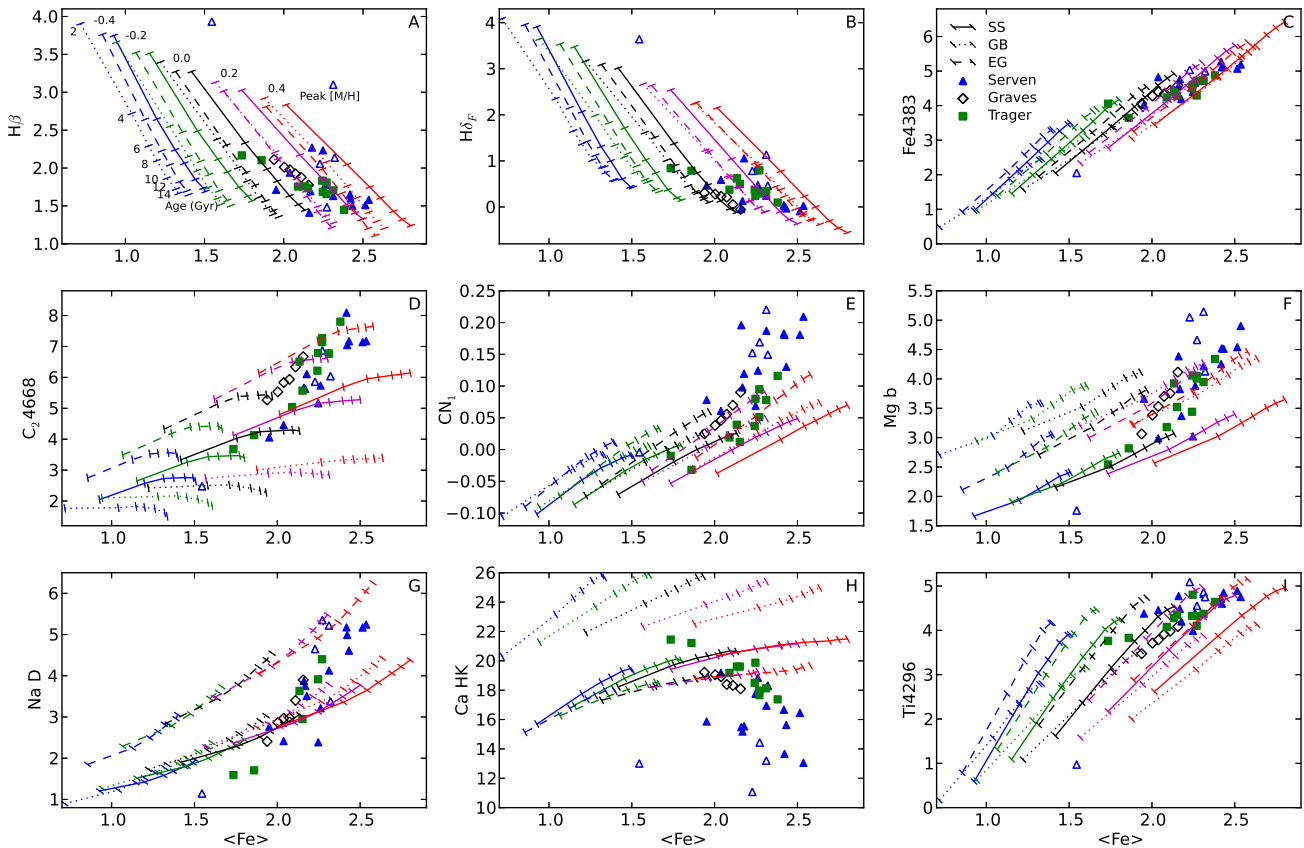} 
\caption{Integrated light index diagrams, all versus $<$Fe$>$. CSP
  models with ADF peaks at [M/H] $=$ -0.4, -0.2, 0, 0.2, 0.4 dex
  (blue, green, black, magenta and red lines, respectively) are shown
  for three chemical mixtures, SS CSPs (solid lines), GB CSPs (dotted
  lines), and EG CSPs (dashed lines). Ages (2 through 14 Gyr in steps
  of 2 Gyr) are shown as ticks along the lines. Serven (blue
  triangles), Graves (open diamonds), and Trager (green squares)
  observations are shown. Five Balmer emission galaxies in the Serven
  sample (open blue triangles) have had additional corrections
  applied.}\label{fig:chem}
\end{figure*}

\textbf{Observational material:} Three sources are used for galaxy
indices in Figure \ref{fig:chem}, which displays how well a bulge
template matches elliptical galaxy observations.

\begin{enumerate}
\item{\citet{gr07} presented an analysis of red sequence galaxy
  spectra ($0.06 < z < 0.08$) from the SDSS, which uses a dual-fiber
  spectrograph, with resolution $R\approx 1800$, wavelength coverage
  $\lambda=3800-9200$\AA, and fiber diameter $d=3''$ which translates
  to physical scales between 3.4 and 4.6 kpc. They labeled the red
  sequence galaxies with neither H$\alpha$ or \oii~detected (at the
  2$\sigma$ level) as ``quiescent'' galaxies, and divided these
  quiescent galaxies (N=2000) into 6 bins in velocity dispersion
  ($\sigma=70-120, 120-145, 145-165, 165-190, 190-220,$ and $220-300$
  km s$^{-1}$). All the spectra were broadened to $\sigma=300$ km
  s$^{-1}$ before stacking. The indices extracted from the stacked
  spectra are represented by open diamonds in Figure \ref{fig:chem}.}
\item{\citet{t08} obtained the spectra of 12 early-type galaxies in
    the Coma cluster with $41<\sigma<270$ km s$^{-1}$, including the
    cD galaxy NGC 4874. The slitlet spectra were corrected along the
    slit to mimic a circular aperture of $2''.7$, which corresponds to a
    physical diameter of 637 pc. These spectra were smoothed to $\sigma=300$ km
    s$^{-1}$. Indices
    are shown as filled squares in Figure \ref{fig:chem}.}
\item{\citet{s10} presented spectra of mostly Virgo elliptical galaxies
  with $80<\sigma<360$ km s$^{-1}$. The observations were taken by the
  Cassegrain Spectrograph mounted on the 4m Mayall telescope at Kitt
  Peak National Observatory. Serven extracted the spectra at an
  aperture of $13''.8$, which corresponds to a physical diameter of
  1.1 kpc for most of the galaxies (the ones in the Virgo cluster),
  and then smoothed the spectra to $\sigma=300$ km s$^{-1}$ .This
  sample is denoted by blue triangles in Figure \ref{fig:chem}.}
\end{enumerate}

\textbf{Age and Metallicity: } Balmer line index -- iron index plots
are known to partially break the age-metallicity degeneracy that
otherwise prevails in most diagrams
\citep{w94a}. $<$Fe$>$\footnote{$<$Fe$>$=(Fe5270 + Fe5335)/2.}
\citep{g93} is sensitive mainly to elemental Fe and thus is a good
analog of [Fe/H]. H$\beta$ is widely used as an age indicator due to
its nonlinear response to main-sequence turnoff temperature. Compared
to bluer H$\delta_F$, H$\beta$ is more susceptible to nebular emission
contamination. We apply nebular emission corrections for H$\beta$ and
H$\delta_F$ to the Graves sample and Serven sample following the
recipe of \citet{sw10}. Five emission-corrected galaxies in the Serven
sample are labelled as open triangles, since larger uncertainties
might be expected for these galaxies in plots involving Balmer
features.  Figures \ref{fig:chem}A and \ref{fig:chem}B show that the
inferred ages are consistent with the expectation that elliptical
galaxies are generally old and metal rich.  One obvious exception is
found at the top of Figs. \ref{fig:chem}A and \ref{fig:chem}B: NGC
3156, a small elliptical galaxy with clear signs of a recent burst of
star formation \citep{kun06}.

All age-sensitive diagrams agree that elliptical galaxies are slightly
more metal rich than the GB, since we model the GB with a peak
metallicity of solar, while the elliptical galaxies in
Fig. \ref{fig:chem} cluster around the $+0.2$ peak line. 

\textbf{Iron:} Figure \ref{fig:chem}C, the Fe4383 -- $<$Fe$>$ plot, is
highly degenerate, showing similar effects of target elements on
Fe4383 and $<$Fe$>$; that is, both indices are sensitive to Fe, age,
and overall abundance at approximately the same rate.  Compared with
SS CSPs, GB CSPs and EG CSPs shift toward weaker values of $<$Fe$>$ by
about 0.1 \AA . This is an artifact of the fact that [M/H] is held
constant, so that enhancement of the target elements comes at the
expense of Fe (c.f., the analytical derivation in \citet{j12}).
Restricted to Fe-sensitive indices, SS, GB, or EG mixtures are all
able to reproduce the observations.

\textbf{Carbon and Nitrogen:} In the C$_2$4668 -- $<$Fe$>$ plot
(Fig. \ref{fig:chem}D), EG CSPs shift upward, but GB CSPs shift
downward with respect to SS CSPs. This is because C$_2$4668 is mainly
controlled by C$_2$ with additional contributions from Fe, Mg, Cr, and
Ti \citep{w94b} plus the evident fact that [C/Fe]$\sim -0.2$ for the GB
but [C/Fe]$=0.21$ for EG. The observations lie close to the EG CSPs,
but the GB is apparently not a good template for this spectral region.

Regarding Fig. \ref{fig:chem}E, CN$_1$ is known to closely depend on
C, N, and O \citep{s05}. The fact that GB CSPs and EG CSPs overlap
means that the amalgamated effects wrought by C, N, and O are almost
the same in these two CSPs, despite the individual abundance ratios
being different ([C/Fe]$\sim -0.2$, [N/Fe]$\sim0.3$, [O/Fe]$\sim0.3$ for
GB\footnote{From $\S$ \ref{sect:SSP}}, [C/Fe]$=0.21$ [N/Fe]$=0.27$,
[O/Fe]$=0.28$ for EG). In Fig. \ref{fig:chem}E, the GB and EG CSPs
shift upward, showing somewhat better agreement with the data points
than SS CSPs.  Some of the data points, especially the Serven sample,
are stronger-lined than the model grid. Partly, that is an artifact of
taking an average value for the models, but it might also be because
the apertures of the Serven spectra concentrate on the nuclei of the
galaxies, zeroing in on the most extremely metal rich stellar
populations and also magnifying near-nuclear effects such as lingering
star formation or low level Active Galactic Nuclei (AGN) activity.

\textbf{Magnesium:} Mg, one of the primary alpha elements, is observed
to be enhanced in massive elliptical galaxies \citep{w92,c14} and its
strong absorption feature at 5170\AA\ is found to be
closely related to galactic velocity dispersion
\citep{bur84,davies93,bbf93, t00,sb06a}. In Figure \ref{fig:chem}F, GB
CSPs and EG CSPs show a much better match to the data points than SS
CSPs. The stellar observations in the bulge and elliptical galaxies
imply [Mg/Fe]$\sim0.21$, and this appears an adequate match to the
data.

\textbf{Sodium:} Na D, a strong absorption feature in the optical, has complex
contributing factors. Na D is very sensitive to the Na abundance and
is somewhat sensitive to the initial mass function (IMF), but it also
suffers from possible interstellar absorption \citep{w94b}.  In Figure
\ref{fig:chem}G, we find that the locus of GB CSPs is not
sufficient to match all the data points. On the other hand, the EG CSP
value of [Na/Fe]$=0.43$ seems to miss the average and match only the
most Na-strong elliptical galaxy data points. In our models, the IMF
is set to a \citeauthor{sal} one, and although IMF variation is not
the subject of this paper, model experiments show that IMF variation
causes only minor changes in the Na D index, leading \citet{j13}, for
example, to conclude that the Na abundance in Na excess
objects might be truly enhanced.

\textbf{Calcium:} Ca HK, first defined in \citet{s05}, was employed to
study the Ca abundance in \citet{w11} in part due to its insensitivity
to the IMF. Given that Ca is one of the alpha elements and Ca yields
closely track Mg \citep{n06,k06}, the decreasing trend of Ca HK as a
function of elliptical galaxy velocity dispersion is
puzzling. \citet{w11} nevertheless concluded that chemical abundance
variation is the explanation for the unusual Ca HK behavior. In Figure
\ref{fig:chem}H we see that enhanced [Ca/Fe], as in the GB CSPs,
drives the model grids even further away from the data points. When we
look at the EG CSPs, [Ca/Fe]$=0.02$, and we would expect almost no
change in Ca HK, but because other element ratios are changing,
especially Mg, the model Ca HK index drops lower. However, even the
amalgamated effects of all the target elements are insufficient to
cover the data points. The problem is eased when one notes that the
Serven sample might lie low due to a spectral response systematic
\citep{w11}.

\textbf{Titanium:} Ti, the heaviest alpha element, is found to display
little evidence of varying as a function of velocity dispersion in
elliptical galaxies \citep{c14,w14}. Here, we estimate Ti abundance by
adopting a rarely used Ti index, Ti4296. The Ti4296 -- $<$Fe$>$ plot
looks similar to a inverted Balmer lines -- $<$Fe$>$ plot due to its
red pseudocontinuum passband encroaching on H$\gamma$, and therefore
has decent capability of breaking the age--metallicity degeneracy.  In
addition, it has relatively low response to most of the heavy elements
\citep{s05}. Figure \ref{fig:chem}I shows that a scaled solar or
mildly enhanced Ti abundance is sufficient for elliptical galaxies.

\section{Discussion}
\label{sect:disc}

The clear basic result is that the [M/H]-dependent elemental mixture
in the GB is not a good template for massive elliptical galaxies,
although the conclusion of \citet{tfw} that the GB is a better
template than local stars also seems perfectly valid. We now go on to
discuss some extensions and implications of this work.

\subsection{Varying the Widths of the ADFs: Red Lean and Red Spread}

The width of the ADF is a parameter with significant importance in our
CSP models. To measure this width via observed spectra or
models is an attractive goal.
Some basic properties of the CSP models
are summarized in Table \ref{tab2}, where we sample only ages 2 and 12
Gyr, and use ADFs that peak at the solar value. The widths (FWHM) of
narrow, normal, wide ADFs are 0.41, 0.62, and 0.93 dex,
respectively. Note that our wide-width ADF is almost the same width as
the classical Simple model (FWHM 1.06 dex), but has a sharper cutoff
at high metal abundance. The weighted means are calculated in the usual way.
For example, to compute the $B$ flux weighted mean metallicity we
convert to a pseudoflux and weight as:
\[ \overline{[M/H]} = {{\sum [M/H]_i w_i }\over{ \sum w_i}} \]
where $i$ is the index over bins in the ADF, $w_i = S_i 10^{-0.4 B_i}$,
$S_i$ is the ADF itself, i.e., the mass fraction of each stellar
population, and $B_i$ is the absolute magnitude of the stellar
population at fixed initial mass.

\begin{table}
\caption{\label{tab2} Mean [M/H] for Composite Populations Peaking at [M/H] = 0 }
\begin{center}
\begin{tabular}{lccc}
\hline \hline
Model & Mass-Weighted & $B$-Weighted & $K$-Weighted  \\
         & [M/H]         & [M/H]    & [M/H] \\
\hline
Narrow, Age 2  & $-0.13$ & $-0.18$ & $-0.15$ \\
Narrow, Age 12 & $-0.13$ & $-0.20$ & $-0.13$ \\
Normal, Age 2  & $-0.19$ & $-0.29$ & $-0.21$ \\
Normal, Age 12 & $-0.19$ & $-0.31$ & $-0.18$ \\
Wide, Age 2    & $-0.26$ & $-0.42$ & $-0.28$ \\
Wide, Age 12   & $-0.26$ & $-0.46$ & $-0.25$ \\
Simple, Age 2  & $-0.25$ & $-0.42$ & $-0.28$ \\
Simple, Age 12 & $-0.25$ & $-0.46$ & $-0.23$ \\
\hline
\end{tabular}
\end{center}
\end{table}

Examination of Table \ref{tab2} tells us that the mean
metallicity is less than the peak metallicity due to the asymmetry of
the ADF. Also, the $B$ band
light samples the ADF at about 0.2 dex lower metallicity than the
mass-weighted or $K$-band-weighted would. This is due to the fact that
metal-poor populations are brighter, coupled with the fact that they
are also bluer. On the other hand, $K$ band light balances the effect of
brighter (even at $K$ band) metal poor populations with the color
change that boosts the $K$ output of metal-rich populations so that
the net effect at $K$ is almost the same as the true, mass-weighted
mean. 

Note that the CSPs investigated in $\S$ \ref{sect:comp} are modelled
with the normal-width ADF.

\begin{figure}
\centering
\includegraphics [scale=0.5]{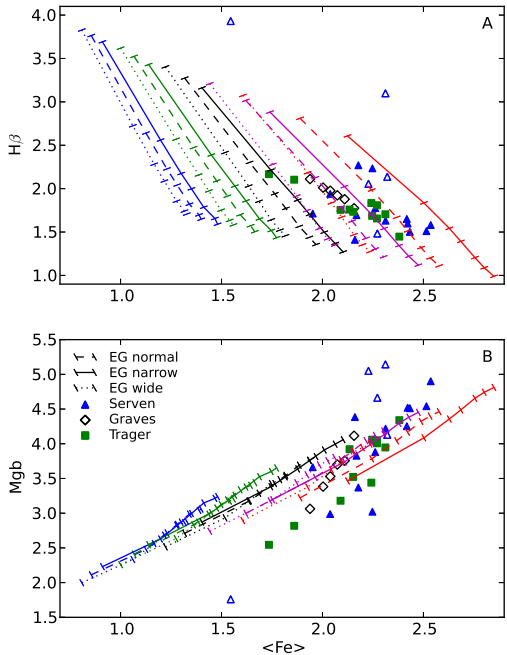} 
\caption{H$\beta$$-<$Fe$>$ and Mgb$-<$Fe$>$ plot for narrow (solid
  lines), normal (dashed lines), and wide (dotted lines) ADFs. Peak [M/H] $=$ -0.4,
  -0.2, 0, 0.2, 0.4 dex are indicated by blue, green, black, magenta and
  red lines, respectively.
  We choose EG CSPs for illustration purpose, but other CSPs would not
  change our conclusions. Elliptical galaxies are labelled the
  same as Figure \ref{fig:chem}.}\label{fig:nnw} 
\end{figure}

Figure \ref{fig:nnw} quantifies the influence of ADF width.  Using the
EG chemical mixture, we plot narrow width (solid), normal (dashed),
and wide (dotted) lines in the H$\beta- <$Fe$>$ and Mgb$-<$Fe$>$
plots. The three elliptical galaxy samples from $\S$ \ref{sect:comp}
are plotted along with the CSP model grids.

We see two main effects. The first effect is that the narrower the
ADF, the more metal-rich it appears (``red lean''). The
second effect is an amplification of the disparity between narrow and
wide ADFs as the peak metallicity shifts to higher [M/H] (``red spread'').

Both red lean and red spread originate from the fact that metal poor
populations are more luminous than their metal rich counterparts (for
the same initial mass). As the width decreases, the more-potent
metal-poor fraction decreases, and the overall average moves toward
the metal-rich in a light-weighted sense as seen in Table \ref{tab2},
or, in looser words, it leans toward the red. The red spread effect
relies upon, in addition, the increased volatility of red giant branch
temperatures for cooler, more metal rich stellar populations (cf. Fig
4 of \citet{w94a}) coupled with the backwarming of cooler stars due to
increased line-blanketing \citep{mihalas70}. The increased $\Delta T$
and color change from backwarming combine to create a greater spectral
change in the metal rich regime than the metal poor regime for the
same $\Delta$[M/H], and thus an amplified spectral change for the more
metal-rich ADFs.

Not all the ADF widths are physically reasonable. With logic similar
to that surrounding Figure 2 of \citet{wdj96} concerning the Simple
Model, the wide-width ADFs seem ruled out by inspection of Figure
\ref{fig:nnw}. That is, in order to make a wide-width model match the
observed spectral indices of elliptical galaxies, the maximum [M/H]
required becomes unreasonably large: Since the peak
= +0.4 model still falls short of observation, a wide-width model that
matches would have \textit{most} of its stars at [M/H] greater than
several times the solar abundance. Although the normal-width ADF (and
not the narrow width) fit nicely, we cannot (yet) show
evidence from integrated light that would exclude the
narrow width ADF, even though star-by-star analyses always show ADFs
wider than our narrow model. 

\subsection{Comparing Chemical Compositions of the Milky Way Bulge and
  Elliptical Galaxies }
\label{sect:diff}

Throughout this paper, we investigate the possibility of using Milky
Way bulge abundance trends to interpret elliptical galaxy
observables. What we found in $\S$ \ref{sect:comp} can be summarized
as this: Fe, Mg, and Ti match the elliptical galaxy abundances, but more C, Na
and less Ca are required to explain the absorption indices of
elliptical galaxies. In other words, MW bulge mimic the
elliptical galaxies better than scaled solar stars do, but concrete
disparities still exist between the bulge and elliptical galaxies.

The MW bulge and elliptical galaxies both consist of old stellar
populations, but they are otherwise dissimilar. The obvious difference
is that the MW bulge is a subcomponent of a spiral galaxy. Also,
\citet{tremaine2002} found MW bulge velocity dispersion is close
to 95 km s$^{-1}$, while the mean velocity dispersion of our
elliptical galaxies is about 200 km s$^{-1}$. The various
nucleosynthetic processes associated with different mass stars are
likely the key to understand the element abundance trends in both
environments, but we still lack a clear connection between galaxy mass
or formation history and nucleosynthetic outcome.

We find more C in elliptical galaxies than in the MW bulge using
C$_2$4668. The N and O abundance can be estimated by including other
element-sensitive indices, like CN$_1$ and TiO$_2$
\citep{gr07,j12,c14}, but since such precise abundance calculation is
beyond the scope of this work, we defer the discussion of N and O
abundances to other papers such as \citet{w14}. In terms of C
enhancement, many studies show that intermediate mass stars contribute
significant amounts of C due to dredge up in the asymptotic giant
branch phase \citep{vg97,woo02}. But \citet{pm04} found the C
abundance generated by these models is lower than the observed
values. Recently, Geneva group \citep{ek12,georgy13} showed rotation
in massive stars significantly changes the C yield. In that spirit,
\citet{pipi09} included stellar rotation in their chemical evolution
models and found an increased C abundance which produce a trend
consistent with \citet{gr07}. It is possible that the discrepancy of C
abundance between the MW bulge and elliptical galaxies might be also
caused by stellar rotation. Following this logic, then for more C we
would require more massive stars to form during the evolution of
elliptical galaxies, since rotation is more significant for massive
stars \citep{ek12, georgy13}.  This could mean a top-heavy IMF, but
not necessarily because increased effectiveness of massive star yields
can be achieved by a \citeauthor{sal} IMF with a galaxy mass-dependent
star formation rate \citep{pipi09}.

The case of calcium is even more puzzling. Given the similarity of old
stellar populations in the MW bulge and elliptical galaxies, the
dichotomous behavior of Ca challenge most of the nucleosynthesis
theories. To summarize the empirical evidence: 1. Ca has sub-solar
abundance in elliptical galaxies. That Ca feature strengths generally
decline with elliptical galaxy velocity dispersion \citep{sag02} is
not a low mass stellar IMF effect \citep{w11} but instead a real
decline in [Ca/Mg], a modest decline in [Ca/Fe], and possibly a small
decline in [Ca/H] as well.  Near solar or sub-solar Ca abundances are
also found by \citet{gr07,j12,w14}. 2. Ca loosely resembles Mg and O
in the MW bulge ($\S$ \ref{sect:cc} \& \citealt{f07}), which seems
agree with its alpha element origin, nevertheless Ca in elliptical
galaxies approximately follows [Fe/H] \citep{gr07,j12,w14,c14}.
According to the Type II supernova yields of \citet{n06}, Ca and Mg
seem in lockstep with each other, trending the same with progenitor
mass and metallicity, and leaving a clear impression that Ca should
follow Mg always. So what is the reason for this Ca-Mg
dissimilarity of the elliptical galaxies?  Two scenarios have been
proposed to explain Ca's atypical alpha element behavior under the
classic one zone, two sources (Type Ia and II supernova) hypothesis.

\begin{enumerate}

\item{\citet{pm04, pipi09} suggested the Ca underabundance relative to
  Mg found in their chemical evolution models is caused by a
  non-negligible Ca contribution via Type Ia supernova
  \citep{n97}. According to the delayed detonation model described in
  \citet{n97}, Ca yield does come out higher than Mg yield for Type Ia
  supernova.  Following this logic, the enhanced Mg abundance in
  massive elliptical galaxies would still come from increased Type II
  contributions, but Ca would be diluted as more Type II products are
  added. Note that is necessary for Type II supernovae to produce less
  Ca than current yields predict for this scheme to be successful. 

More speculatively, there
  might be a connection with the recent studies of silicon group 
  elements: Based on the classical W7 models
  \citep{nomoto84,thiele86}, \citet{de14} propose the ``W7-like''
  models that is now capable of varying the $^{22}$Ne mass fraction
  freely.  They notice the electron fraction of the progenitor white
  dwarfs, which anti-correlates with the $^{22}$Ne mass fraction,
  systematically influences the nucleosynthesis of the silicon group
  elements (Si, S, and Ca) in the sense that Ca shows a nearly
  quadratic increasing trend with electron fraction, while Si is
  almost insensitive to the electron fraction. Observationally
  speaking, Si is a fair alpha element compared to Ca, since it
  roughly follows Mg and O in both the MW bulge and elliptical
  galaxies\footnote{S is left out due to a lack of spectral lines
    \citep{s05}.}. Therefore, the observations imply more Ca, but the
  same amount of Si are generated in low mass ellipticals via Type Ia
  supernova. The Ca and Si yields would reconcile with the theory of
  \citet{de14} if the electron fraction is a decreasing function of
  stellar velocity dispersion. However, we also notice the dynamical
  range of electron fraction is small in \citet{de14}, which means an
  observational demonstration is unlikely.

If the MW bulge is an analog of a low-$\sigma$ elliptical galaxy
\citep{mac09}, its Ca contribution from Type Ia supernova should be 
higher than massive elliptical galaxies (c.f. \citealt{tom11}
for a similar statement). Given that the decreasing trend of
[$\alpha$/Fe] versus [Fe/H] is usually explained by a Type Ia
supernova contribution of Fe \citep{wst1989}, an increase of Type Ia
supernova Ca would flatten its negative slope against
[Fe/H], and make Ca dissimilar to other alpha
elements. Unfortunately for the Type Ia Ca origin hypothesis, the [Ca/Fe]
and [Mg/Fe] trends in Figure \ref{fig:elem} look very similar. } 

\item{Alternatively, the Ca behavior could be conceived to be the
  result of Ca yield suppression in massive ellipticals. This
  conjecture might be caused by a mass-dependent Ca yield in
  supernova, where the highest-mass stars contribute less Ca but
  continue to contribute Si, O, and Mg, and massive ellipticals would
  favor these highest-mass stars. This was suggested long ago
  \citep{w92} but the specific question of Ca yield is ambiguous
  \citep{ww95,n06}. On the other hand, Ca yield suppression might
  connect with a metallicity-dependent
  yield. \citet{f07} suggested inclusion of metallicity-dependent
  wind may change the alpha element yields from massive Type II
  progenitors in \citet{ww95}. As metallicity increases, the yield
  of hydrostatic elements (e.g., O, Mg) increases and the yield 
  of explosive elements (e.g., Si, Ca, Ti) declines. This seems
  qualitatively correct.
  A metallicity-dependent
  yield has not been systematically studied by any
  theoretical group to the best of our knowledge. }
\end{enumerate}

\subsection{Recovering Abundances with Simple Stellar Population Models}
\label{sect:SSP}

Efforts to discover the detailed chemical composition of galaxies and
most efforts to discover the ages of old stellar populations use
single-burst, single-composition simple stellar population (SSP)
models. Going from inherently CSP populations but interpreting them
via SSP models might introduce systematics. To explore this, we
perform the following theoretical exercise.

In \citet{w14}, a chemical mix-sensitive inversion program that uses
SSP based on \citet{b94} evolution was applied to several samples of
galaxies. Here, we use the same inversion program on the CSP models,
built with evolution based upon \citet{mar08} as well as being
composite in metallicity. Uncertainties in each index were calculated
by a photon noise model normalized to S/N = 25 per \AA\ at 5000\AA\ in
$F_\lambda$ units. The comparison is presented in Table \ref{tab1} for
the average of two ages (8 and 12 Gyr) computed from a CSP model with
normal-width ADF peaked at solar abundance and with a run of bulge
chemical mixtures and more recent stellar evolution, interpreted with
older models under a single-burst hypothesis. As in \citet{w14},
sensitivities to different elements were turned on and off in order to
generate a crude ``permutational'' estimate of uncertainty.

\begin{table}
\caption{\label{tab1} Recovery of Population Parameters Under an SSP Hypothesis}
\begin{center}
\begin{tabular}{cccc}
\hline \hline
Quantity & Mass-Weighted & Recovered & Standard  \\
         & Value         & Value     & Deviation \\
\hline
Age       & 8       &  5.7 & 0.9 \\
Age       & 12      &  10.6 & 2.2 \\
{[M/H]}   & $-0.19$ & $-0.19$ & $0.15$ \\
{[C/Fe]}  & $-0.23$ & $-0.27$ & 0.05 \\
{[N/Fe]}  & 0.38    & 0.14 & 0.09 \\
{[O/Fe]}  & 0.32    & 0.24 & 0.19 \\
{[Na/Fe]} & 0.06    & 0.17 & 0.17 \\
{[Mg/Fe]} & 0.25    & 0.25 & 0.14 \\
{[Si/Fe]} & 0.17    & 0.25 & 0.03 \\
{[Ca/Fe]} & 0.19    & 0.20 & 0.07 \\
{[Ti/Fe]} & 0.18    & 0.37 & 0.18  \\
\hline
\end{tabular}
\end{center}
\end{table}

Table \ref{tab1} shows approximate agreement between CSP input and
SSP-recovered parameters, with interesting systematic drifts. The mean
ages drift slightly younger, especially with [Si/Fe] set to
zero. Letting [Si/Fe] vary allows the ages to relax about 1.5 Gyr
older. This is due to the molecular SiH features in the blue covarying
with some of the Balmer indices, from which the bulk of the age
information comes. With some permutations of which elements were
allowed to vary, there arises a correlation between age and a few of
the [X/Fe], often [Na/Fe] for example. This is slippage in the
inversion scheme and a caution for future work with this inversion
program. The mean [M/H] is recovered well.  The abundance parameters
are adequately recovered, with [N/Fe] too low, and [Ti, Si/Fe] too
high at marginal statistical significance.  Table \ref{tab1} is
encouraging in terms of basic coherence between models and in the fact
that the element ratios can be recovered in a way that resembles the
actual, metallicity-and-mass-weighted mean. One can also safely say
that more work is needed to truly understand all possible systematics.

\subsection{Detectability of ADF Width}

\begin{figure}
\hspace*{-0.15in}
\includegraphics [scale=0.45]{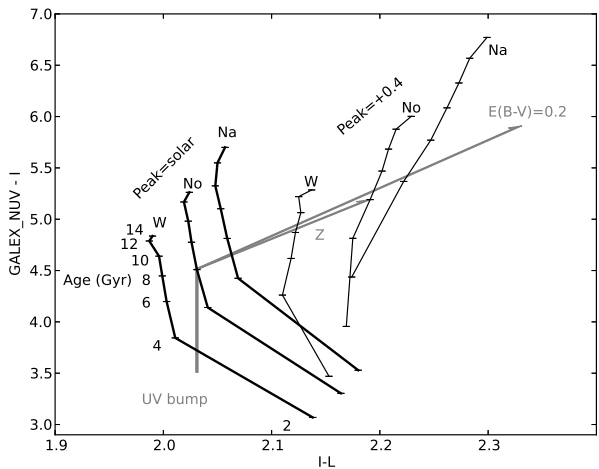} 
\caption{Integrated light color-color diagram that shows sensitivity
  to the width of the ADF involving the GALEX NUV filter (0.2 $\mu$m)
  and Johnson-Cousins $I$ (0.7 $\mu$m) and $L$ (3.4 $\mu$m). Colors
  are set to zero magnitude for the spectral shape of Vega. As in
  previous figures, stellar population ages between 2 and 14 Gyr are
  shown, for three ADF widths (marked ``W'' for wide, ``No'' for
  normal, and ``Na'' for narrow) and two peak metallicities, peak
  [M/H] = solar (bold lines) and peak [M/H] = +0.4 (thinner
  lines). The effect of dust screen extinction, the increase of
  metallicity, and including a hot UV-bump component are illustrated
  with labelled vectors.}\label{fig:detect}
\end{figure}

We are able to fuzzily reject a too-wide abundance distribution, but
we are so far unable to distinguish narrow from normal ADFs in
integrated light. We now attempt to point a way forward by considering
the spectral shape of the composite population as a
whole. Compositeness of stellar populations, be it age-composite or
abundance-composite, can be expected to increase the variety of
stellar temperatures present in the population, and thus increase the
width of the overall integrated sum of near-blackbody stellar
fluxes. Thus, sampling the spectrum over a broad span of wavelength
may yield diagnostic information. We attempt to illustrate how this
might work in Figure \ref{fig:detect}.

We note several encouraging things. In that color plane, age and
abundance effects are nicely orthogonal. The effects of ADF width
increase as the population ages. If we measure UV bump strength from
farther in the UV and if we measure the metallicity from optical
spectral indices, then it does indeed seem as if the narrow and normal
widths can be distinguished, with the only effect unaccounted for
being dust extinction. We therefore tentatively judge that the
prospects are bright for measuring the ADF width galaxies whose ages
are fairly unimodal and whose dust content is characterized
well. Using near-UV spectral indices as a proxy for photometric color
would presumably lessen the effects of dust extinction and make the
method more robust.

\section{Summary}
\label{sect:con}

Using the evolving Worthey models \citep{w94a,t08,lee09,tw13,lickx},
we investigate composite stellar population models with different
metallicity spreads. In integrated light, the narrower ADF appears
more metal rich (``red lean'') and the disparity between narrow and
wide ADFs widens as peak [M/H] move to a higher metallicity (``red
spread''). Wide-width ADFs and the Simple Model ADF are ruled out
because of the need for unreasonably large [M/H]\footnote{For
  additional discussion of the Simple Model ADF, see
  \citet{wdj96}}. Star by star, we confirm that the normal-width ADF replicates the
observed solar neighborhood and MW bulge ADFs, and our modeling 
indicates that the normal-width or narrow-width ADF also fits 
elliptical galaxies in integrated light.

The good match of elliptical galaxy colors and TiO strengths with MW
bulge templates \citep{tfw} inspires us to find out if the MW bulge
element abundance trend can be applied to elliptical galaxies. We
model CSPs with MW bulge chemical compositions, tracking the detailed
behavior of each element as a function of [M/H], CSPs with scaled
solar abundances, and CSPs with an elemental mixture chosen to match
massive elliptical galaxies \citep{c14}, then compare model absorption
feature indices with observed indices. Iron, Ti, and Mg vector about
the same for the MW bulge and elliptical galaxies, while the trends of
C, Na, and Ca are different. If the MW bulge is analogous to a low
velocity dispersion elliptical galaxy, attempting to discern
implications about elliptical galaxy evolution based on various
nucleosynthetic processes associated with different environments (star
formation timescales, galactic winds, stellar population IMF,
supernova mass- and metallicity dependent yields, and white dwarf
electron density) leads to no clear astrophysical frontrunner as to
the cause of the abundance trends.

We perform an exercise to feed our MW bulge CSP models based on one
set of  stellar evolutionary isochrones into a chemical
mixture inversion program based on a different set of stellar
evolutionary isochrones and find that the parameters were recovered
fairly well in the sense of matching the ADF-weighted mass average
abundances. Our particular exercise uncovered a systematic trend
toward younger ages. Elemental mixtures are recovered well, though in our case
[N/Fe] came out on the low side, while [Ti/Fe] and [Si/Fe] were too
high.  The age and abundance systematics uncovered serve to urge further
study of the inversion process.

We investigate whether the width of an ADF could be measured from
integrated light alone, and find a successful technique: photometric
colors of very long wavelength span. The main caveats for old stellar
populations are dust content and the correct subtraction of whatever
ultraviolet-bright subpopulation might be present.

\section{acknowledgement} 
The authors wish to thank Washington State University and its
Department of Physics and Astronomy, G. J Graves, S. C. Trager, and
J. Serven for the use of their data, and an anonymous referee for
helpful comments. This research has made use of the VizieR catalogue
access tool, CDS, Strasbourg, France.

\label{lastpage}

\end{document}